\documentclass[prl,superscriptaddress,showpacs,floats,floatfix,preprint]{revtex4}

\usepackage{epsfig}
\begin{document}

\draft

\title{Formation of the Pd atomic chain in hydrogen atmosphere}

\author{Manabu Kiguchi}
\thanks{Present address: Department of Chemistry, Graduate School of Science and Engineering, 
Tokyo Institute of Technology 2-12-1 W4-10 Ookayama, Meguro-ku, Tokyo 152-8551, Japan, E-mail: kiguti@chem.titech.ac.jp}
 \affiliation{Division of Chemistry, Graduate School of Science, Hokkaido University, N10W8, Kita, Sapporo, 060-0810, Japan}

\author{Kunio Hashimoto}
\affiliation{Division of Chemistry, Graduate School of Science, Hokkaido University, N10W8, Kita, Sapporo, 060-0810, Japan}

\author{Yuriko Ono}
\affiliation{Division of Chemistry, Graduate School of Science, Hokkaido University, N10W8, Kita, Sapporo, 060-0810, Japan}

\author{Tetsuya Taketsugu}
\affiliation{Division of Chemistry, Graduate School of Science, Hokkaido University, N10W8, Kita, Sapporo, 060-0810, Japan}

\author{Kei Murakoshi}
\affiliation{Division of Chemistry, Graduate School of Science, Hokkaido University, N10W8, Kita, Sapporo, 060-0810, Japan}

\date{\today}

\begin{abstract}
The formation of a Pd atomic chain in a hydrogen atmosphere was investigated by measurements of 
conductance and vibrational spectroscopy of a single molecular junction, and the theoretical 
calculation. While atomic chains were not formed for clean 3d and 4d metals, in the case of Pd (a 
4d metal) atomic chains could be formed in the presence of hydrogen. Stable atomic chains with 
two different atomic configurations were formed when the Pd atomic contact was stretched in a 
H$_{2}$ atmosphere; highly conductive short hydrogen adsorbed atomic chain and low conductive long 
hydrogen incorporated atomic chain.

\end{abstract}

\maketitle

\section{INTRODUCTION}
\label{sec1}

Chains of single metal atoms forming bridges between metal electrodes have attracted wide 
attention due to their unique properties \cite{2} and their potential applications as ultimate 
conducting wires in nanoscale electronic devices \cite{3}. In the case of Au, atomic chains up to 2 
nm long have been created in UHV, and their physical properties have been investigated with 
various techniques \cite{3,4,5}. The formation of atomic chains was also reported for other 5d 
metals, such as Pt and Ir. On the other hand, it has not been possible to prepare atomic chains in the 
case of 3d and 4d metals for the following reasons \cite{3,6}. The formation of the atomic chain 
requires relatively strong metal-metal bonds within the chain compared to other bonds in regions 
close to the chain. In the case of 5d metals, the bonds in the atomic chain, in which all atoms are at 
the surface, is strengthened due to the relativistic effects of the valence electrons \cite{3,6}. In 
contrast, the stabilization of the bond strength in atomic chain made from 3d and 4d metals is not 
significant, making the fabrication of chains from these metals extremely difficult. Despite these 
difficulties, however, various interesting properties have been predicted. For example, theoretical 
studies have shown that the Pd atomic chain is ferromagnetic, even if the length of the atomic chain 
is short ($\sim$3 atom) \cite{2}. Consequently, successful fabrication of atomic chains of 3d and 4d 
metals, Pd in particular, is of great scientific interest. 

Recently, it was shown that atomic or molecular adsorption on metal surfaces can stabilize the 
metal atomic contact. 2 nm long Ag atomic chains were created in the presence of oxygen in UHV 
at 4 K, while clean Ag only forms short chains \cite{7}. In the case of Pd, Pd atomic chains were 
stabilized under the hydrogen evolution reaction in solution \cite{8}. In UHV, the conductance of 
the Pd atomic contact decreased from 2$G_{0}$ ($G_{0}$  = 2$e^{2}/h$ to 0.5$G_{0}$ due to the 
introduction of hydrogen \cite{9}. While it was revealed that it is possible for the metal atomic 
chain to be stabilized in the presence of other atoms and molecules, the structure of the stabilized 
atomic chain is not clear. It is not known whether a molecule adsorbs on the chain surface or if it is 
incorporated into the chain. Vibrational spectroscopy of a single molecular junction is one 
promising technique used to investigate the atomic configuration of the single molecular junction 
\cite{10,11}. In the present study, we report on the formation of Pd atomic chains in a hydrogen 
atmosphere. The structure of the atomic chain has been verified by comparing conductance 
measurements and vibrational spectra of a single molecular junction to those from theoretical 
calculations. 

\section{EXPERIMENTAL}
\label{sec2}

All measurements were performed using a mechanically controllable break junction (MCBJ) 
technique (see Ref.\cite{3} for a detailed description). A notched Pd (0.1 mm diameter, 
99.99$\%$ purity) wire was glued onto a bending beam and mounted in a three-point bending 
configuration inside a vacuum chamber. Once under vacuum and cooled to 4.2 K, the Pd wire was 
broken by mechanical bending of the substrate. The bending was then relaxed to form atomic-sized 
contacts between the wire ends using a piezo element for fine adjustment. H$_{2}$ was injected 
into the chamber via a capillary. DC two-point voltage-biased conductance measurements were 
performed by applying a voltage in the range from 10 to 150 mV. AC voltage bias conductance 
measurements were performed using a standard lock-in technique. The conductance was recorded 
for a fixed contact configuration using a 1 mV AC modulation at a frequency of 7.777 kHz while 
slowly ramping the DC bias between -100 and +100 mV.

\section{EXPERIMENAL RESULTS}
\label{sec3}
\begin{figure}
\begin{center}
\leavevmode\epsfxsize=80mm \epsfbox{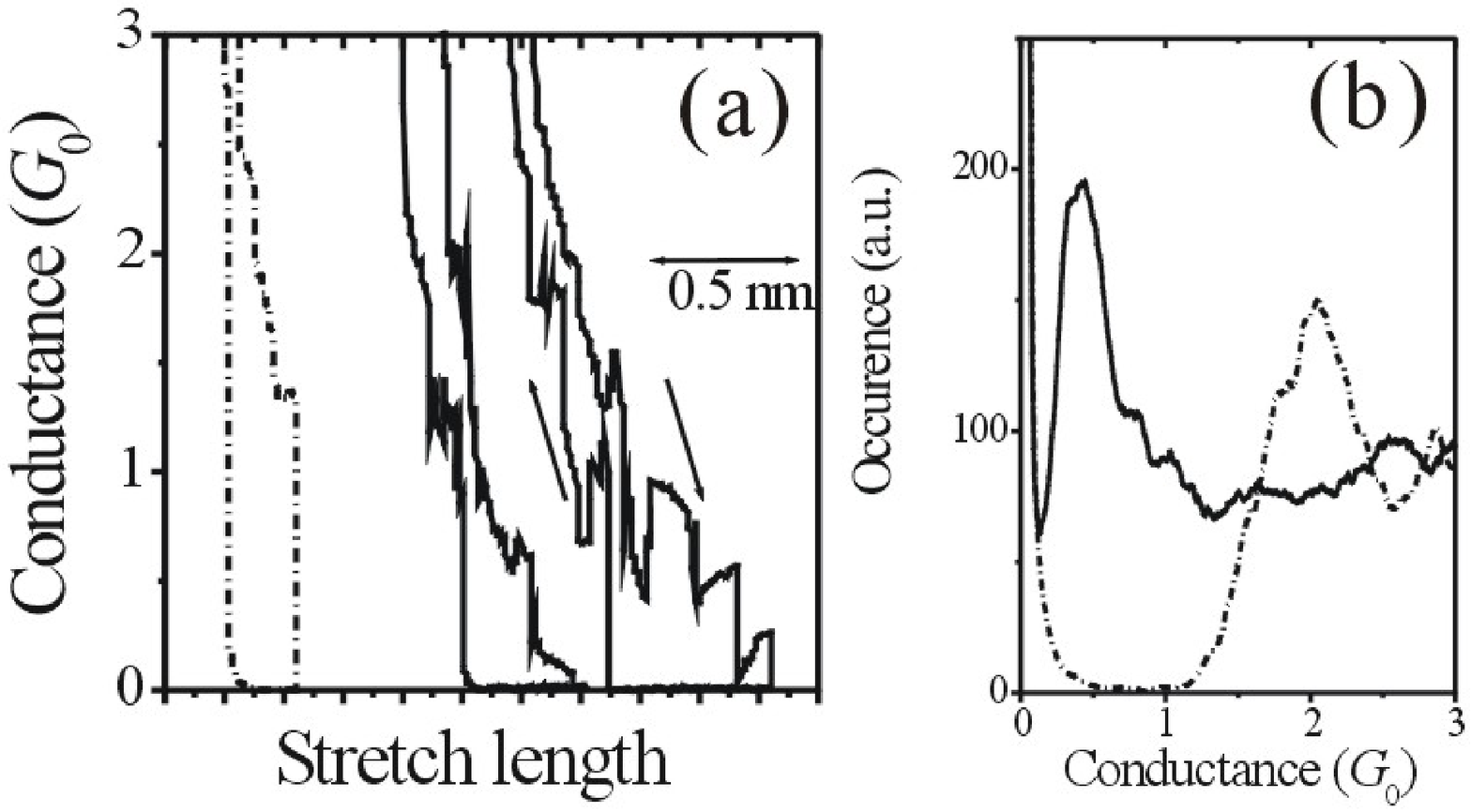}
\caption{(a) Breaking and return conductance traces for clean Pd contacts (dotted line), and for  
Pd contacts in a H$_{2}$ atmosphere (line). (b) Conductance histograms for the clean Pd contacts 
(dotted line), and for Pd contacts in a H$_{2}$ atmosphere (line). Each conductance histogram was 
constructed from 1000 conductance traces recorded with a bias voltage of 0.1 V during the breaking 
of the contact.}
\label{fig1}
\end{center}
\end{figure}

\begin{figure}
\begin{center}
\leavevmode\epsfxsize=80mm \epsfbox{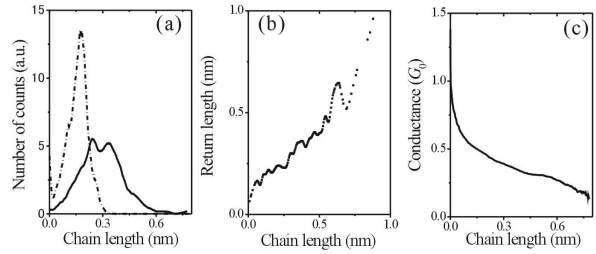}
\caption{(a) Length histogram for clean Pd contacts (dotted line) and Pd contacts in a H$_{2}$ 
atmosphere (line).(b) Average return lengths as a function of chain length. (c) Average 
conductance as a function of chain length.  Each curve in (a)-(c) was constructed from 2000 conductance 
traces recorded with a bias voltage of 0.1 V 
during the breaking of the contact. } 
\label{fig2}
\end{center}
\end{figure}

Figure~\ref{fig1} (a) shows typical conductance traces for clean Pd contacts and for Pd 
contacts in a H$_{2}$ atmosphere, recorded with a bias voltage of 0.1 V. For the clean Pd contacts, 
the conductance decreased in a stepwise fashion and the corresponding conductance histogram (Fig. 
~\ref{fig1}(b)) showed a peak near 2 $G_{0}$, which corresponds to a clean Pd atomic contact. 
After the introduction of H$_{2}$, a plateau below 1 $G_{0}$ was frequently observed and the 
corresponding histogram (Fig.~\ref{fig1} (b)) showed a peak around 0.4 $G_{0}$. The obtained 
conductance histogram agreed with the previously reported results \cite{9}. Here it should be 
noticed that the Pd atomic contact in a H$_{2}$ atmosphere could be stretched to quite long 
lengths ($\sim$0.5 nm), which suggested the formation of an atomic chain. In order to investigate 
the chain formation, the length histogram was investigated for the last plateau. Figure~\ref{fig2}(a) 
shows the length histogram of the last conductance plateau for the Pd contacts in a 
H$_{2}$ atmosphere, together with that for the clean Pd contacts. The length of the last plateau was 
taken as the distance between the points at which the conductance dropped below 1.0 $G_{0}$ and 
0.05 $G_{0}$ for the Pd contact in a H$_{2}$ atmosphere, and the boundaries were 2.5 
$G_{0}$ and 1.0 $G_{0}$ for the clean Pd contact. The length histogram for the Pd contacts in a 
H$_{2}$ atmosphere did not change when the upper boundary was changed from 1.0 $G_{0}$ to 
2.5 $G_{0}$. The length histogram showed that the Pd atomic contact could be stretched as long as 
0.5 nm in a H$_{2}$ atmosphere, while the clean Pd contact broke within 0.3 nm. Since the 
Pd-Pd distance is 0.27 nm for bulk Pd, the final 0.5 nm long plateau corresponds to an atomic chain 
of approximately two Pd atoms. 

The formation of the Pd atomic chain was supported by the analysis of the conductance trace. 
Fig.~\ref{fig2} (b) shows the average return lengths as a function of chain length, recorded with a 
bias voltage of 0.1 V. The return length was defined as the distance over which the two electrodes 
were moved back to make contact after the junction broke. The return length was averaged over 
2000 breaking cycles. Apart from an offset of 0.1 nm due to the elastic response of the banks, the 
average return length was approximately proportional to the chain length, suggesting that a fragile 
structure was formed with a length corresponding to that of the last plateau \cite{3,5,6}. The long 
plateau over 0.5 nm length and the one by one relationship between the chain and return length 
indicated that a Pd atomic chain could be formed in a H$_{2}$ atmosphere. Figure~\ref{fig2}(c) 
shows the average conductance of the atomic chain as a function of the chain length. The curve was 
obtained by adding all measured conductance traces from the start value (1 $G_{0}$) onward, and 
dividing each length by the number of traces included at that point. It was observed that the mean 
conductance of the atomic chain decreased with increasing chain length. 

\begin{figure}
\begin{center}
\leavevmode\epsfxsize=80mm \epsfbox{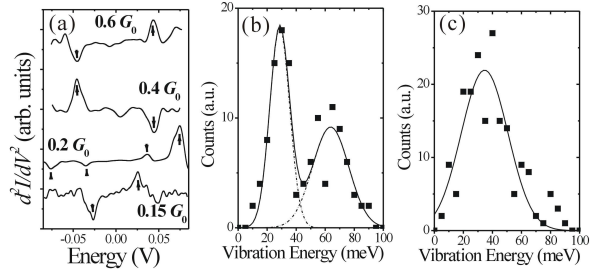}
\caption{(a) Typical $d^{2}I/dV^{2}$ curves for Pd contacts in a H$_{2}$ atmosphere taken at conductance 
ranging from  0.15$\sim$0.6 $G_{0}$. (b) The distribution of vibrational energy for the contacts with conductance 
below 0.3 $G_{0}$. (c) The distribution of vibrational energy for contacts with conductance ranging from  
0.3$\sim$0.6 $G_{0}$.} 
\label{fig3}
\end{center}
\end{figure}

In order to investigate the atomic configuration of the atomic chain, the vibrational 
spectroscopy of a single molecular junction was measured for the Pd atomic contacts in a 
H$_{2}$ atmosphere. Figure~\ref{fig3}(a) shows examples of the $d^{2}I/dV^{2}$ curves 
(vibrational spectroscopy of a single molecular junction) for the Pd atomic contacts in a 
H$_{2}$ atmosphere taken at a zero bias conductance of 0.15$\sim$0.6 $G_{0}$. The spectral 
shape and energy of the vibrational modes varied with the contacts. Peaks and dips were observed 
in the spectra depending on the conductance of the contacts. When the conductance of the contact 
was lower than 0.4 $G_{0}$, only peaks were observed in the spectra. Conversely, both peaks and 
dips were observed in the spectra when the conductance of the contact was larger than 0.4 
$G_{0}$. 

The peaks in the spectra indicated a conductance enhancement above a certain voltage, while 
the dips indicated a conductance suppression. The conductance enhancement can be explained by 
the opening of an additional tunneling channel for electrons that lose energy to vibrational modes 
\cite{10,11,12,13}. Likewise, the conductance suppression could be explained as the 
backscattering of electrons that lose energy to vibrational modes in the ballistic contacts in which 
the electron transmission probability is close to one. When a molecule symmetrically couples to 
both metal electrodes and the number of the conduction channel is one, the theoretical investigation 
predicts that the conductance of the contact would be enhanced by phonon excitation for contacts 
having a conductance below 0.5 $G_{0}$, while the conductance would be suppressed for the 
contacts having a conductance above 0.5 $G_{0}$ \cite{12,13}. The experimental result that only 
peaks were observed for the contacts having conductance below 0.4 $G_{0}$ agrees with this 
theoretical prediction. Furthermore, the results that both peaks and dips were observed for contacts 
with conductances above 0.4 $G_{0}$, indicate the presence of more than one conductance channel. 
In the latter case, the presence of more than one channel makes it difficult to discuss the direction of 
the conductance change induced by the phonon excitation \cite{12,13}. 

Our observations indicate that the Pd atoms bridge between the Pd electrodes for contacts with 
conductances above 0.4 $G_{0}$, as discussed below. If a hydrogen bridges the gap between the 
Pd electrodes, electrons would be transmitted through 1s atomic orbital (in the case of a hydrogen 
atom bridge) or bonding or anti-bonding molecular orbitals (in the case of a hydrogen molecule 
bridge) which would be modulated by Pd. Since these orbitals are not degenerate and are 
energetically well separated, only one orbital can contribute to the electron transport through the 
hydrogen atom or molecule bridge. Since the present experimental results suggested multiple 
conduction channels, hydrogen can not have formed a bridge between the Pd electrodes as electrons 
would have been transmitted through Pd d orbitals.  

In the spectra of Fig.~\ref{fig3}, the vibrational modes were observed at 25 meV 
(conductance of contact: 0.15 $G_{0}$), 35 and 75 meV (0.2 $G_{0}$), 45 meV (0.4 $G_{0}$) 
and 45 meV (0.6 $G_{0}$). While the energy of the vibrational modes varied with the contacts, the 
distribution shows some defining characteristics. Specifically, it depended on the conductance value 
of the contacts. The distribution of the vibrational energy of the contacts having conductance below 
0.3 $G_{0}$ was different from that above 0.3 $G_{0}$. Figure 3(b) and (c) show the distribution 
of vibrational energy for contacts having conductances below 0.3 $G_{0}$ and 0.3$\sim$0.6 
$G_{0}$, respectively. The histograms were obtained from 250 spectra of different contacts. The 
histogram showed two peaks at 29 meV (14 meV in width) and 64 meV (25 meV in width) for 
contacts with conductance below 0.3 $G_{0}$ and broad single peak at 35 meV (32 meV in width) 
for contacts with conductance of 0.3$\sim$0.6 $G_{0}$. Since the conductance and 
vibrational energy of the atomic contacts is strongly correlated with their atomic configurations, the 
present results indicate that atomic chains with two different atomic configurations would be 
formed during stretching of the Pd contacts in a H$_{2}$ atmosphere. The conductance of the 
atomic chain decreased with the chain length, as shown in Fig.~\ref{fig1}(a) and Fig.~\ref{fig2} 
(c). The poorly conductive long atomic chain shows two vibrational modes around 29 and 64 meV, 
while the highly conductive short atomic chain shows one vibrational mode around 35 meV. Here it 
should be noted that the spectra for the contact of 0.2 $G_{0}$ (Fig.~\ref{fig3}(a)) shows two 
vibrational modes, indicating the presences of two vibrational modes in the contact. The double 
peaks in the histogram did not, however, originate from the two atomic contacts with different 
vibrational modes and similar conductance values. Briefly, we comment on the features of the 
conductance histogram. While the vibrational spectra suggest the formation of atomic chains with 
two different atomic configurations, the conductance histogram shows a single feature around 0.4 
$G_{0}$. We think that single feature in the conductance histogram is composed of two broad 
peaks. The formation of atomic chains with two different atomic configurations is supported by the 
conductance trace. The last curve in Fig.~\ref{fig1}(a) shows steps around 0.5 $G_{0}$ and 0.2 $G_{0}$.

\section{CALCULATION RESULTS}
\label{sec4}

To gain insight regarding the structural and vibrational modes of our Pd chain in a 
H$_{2}$ atmosphere, we performed density functional theory (DFT) calculations using the 
Gaussian03 program \cite{14}. The hybrid DFT approach, B3LYP \cite{15}, was employed with 
the Stuttgart-Koln effective core potentials for Pd \cite{16} and Dunning's cc-pVDZ basis set for H 
\cite{17}. The atomic contact part was modeled using two types of clusters, (a) 
Pd$_{4}$-Pd-Pd$_{4}$ and (b) Pd$_{4}$-Pd$_{4}$, where a hydrogen molecule was introduced in the 
area around the central Pd atoms. The geometry was fully optimized for the respective models and 
normal mode analyses were performed for the obtained structures. Through preliminary calculations, 
we have found that H$_{2}$ cannot attach to the central part of Pd chains as a molecule, and easily 
dissociates to hydrogen atoms on the Pd cluster. Thus, geometry optimizations were performed for 
the respective model clusters with the adsorption of two hydrogen atoms and we succeeded in 
locating minimum energy structures with no imaginary frequencies for both model clusters.

\begin{figure}
\begin{center}
\leavevmode\epsfxsize=130mm \epsfbox{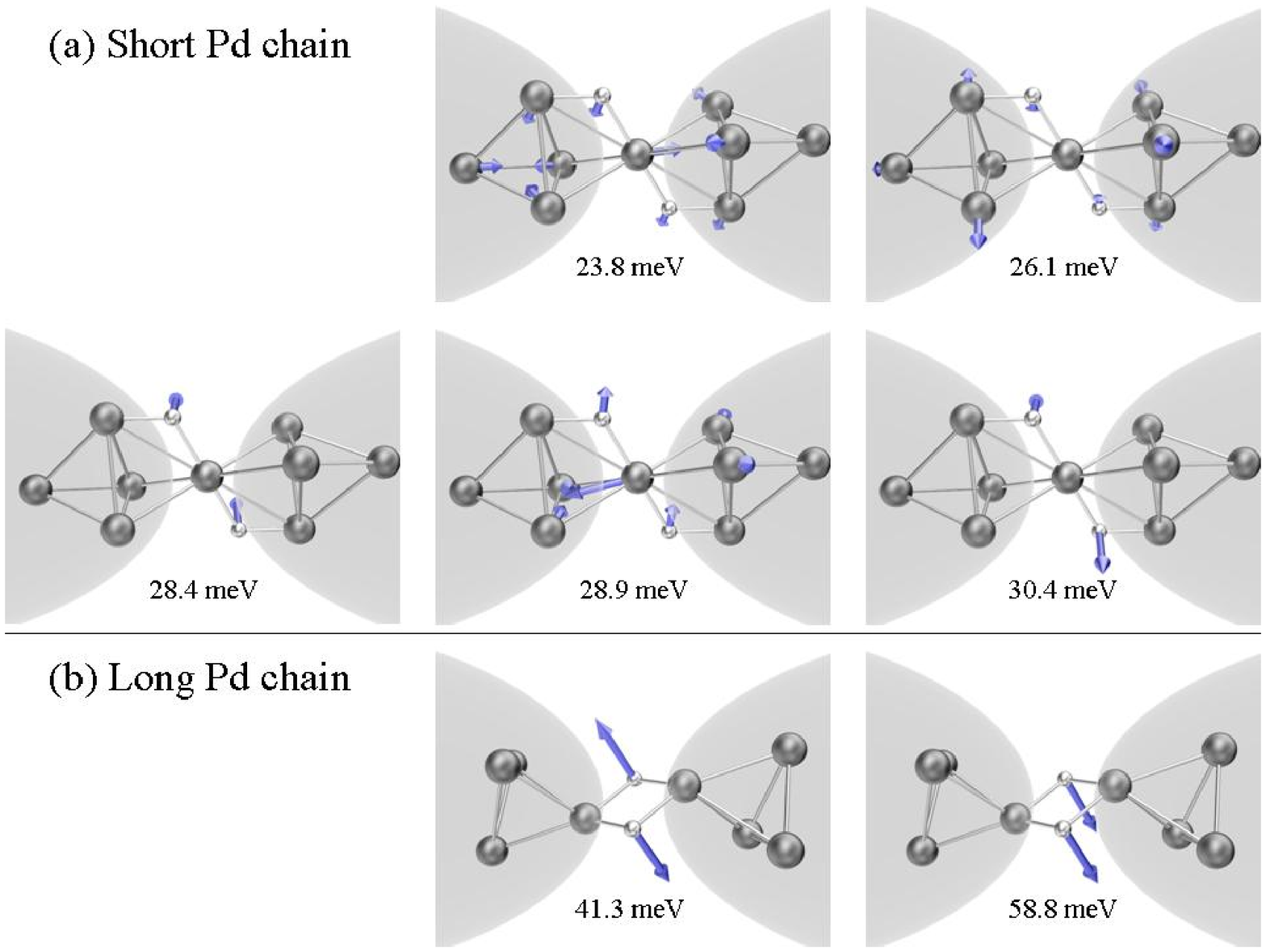}
\caption{Vibrational modes containing the movement of hydrogen atoms calculated for the 
optimized structures of (a) Pd$_{4}$-H-Pd-H-Pd$_{4}$ (short Pd chain) and (b) Pd$_{3}$-Pd-H-H-Pd-Pd$_{3}$ (long Pd 
chain). The length of the arrows indicates the amplitude of the modes.} 
\label{fig4}
\end{center}
\end{figure}

Figure~\ref{fig4} shows the vibrational modes containing the movement of hydrogen atoms 
calculated for the optimized structures of (a) Pd$_{4}$-H-Pd-H-Pd$_{4}$ (referred to as "short Pd chain") 
and (b) Pd$_{3}$-Pd-H-H-Pd-Pd$_{3}$ (referred to as "long Pd chain"). The H$_{2}$ molecule dissociates to H 
atoms in both structures ($r_{HH}$ $\sim$0.192 nm in structure (b)) which work to stabilize the 
contact part of the Pd chain. The present calculations reveal that the short hydrogen-adsorbed shows 
the 5 vibrational modes of the Pd chain around 23-31 meV, while the long hydrogen incorporated 
Pd contact shows the vibrational modes at 41 and 59 meV. 

The vibrational modes observed in the present experiments are discussed based on these 
calculation results and previously reported results of nanocontacts. Pd nanocontact and Pd 
nanocontact in a H$_{2}$ atmosphere were investigated using point contact spectroscopy. In 
these experiments, Pd crystal vibrational modes were observed around 15-20 meV for the Pd 
nanocontact and, likewise, the vibrational modes of dissolved H atoms was observed around 60 
meV for contacts in a H$_{2}$ atmosphere. In the present study, the histogram of the vibrational 
modes showed two peaks around 29 meV and 64 meV for the long atomic chain, and a broad single 
peak around 35 meV for the short atomic chain. Based on these experimental and theoretical 
calculation results, an experimentally obtained broad single peak around 35 meV for the short 
atomic chain would correspond to the 5 calculated vibrational modes ranging from 23-31 meV. 
Since energy differences among the five modes are small, they appear as a single broad peak in the 
histogram. 35 meV is close to the energy of the vibrational mode of the Pd crystal (15-20meV) and 
the vibrational modes of Pd in the stem parts of the electrodes might contribute to the 35 meV peak. 
The two experimentally obtained peaks around 29 meV and 64 meV for the long atomic chains 
correspond to the calculated vibrational modes at 41 meV and 59 meV. For the long atomic chains, 
the vibrational modes of Pd in the stem part of the electrodes and the modes of dissolves H atoms in 
the Pd crystal can contribute to the peaks around 29 meV and 64 meV, respectively. In both short 
and long atomic chains, the peaks in the histogram of the vibrational modes were broad. The atomic 
chains whose atomic configuration was slightly modified from the stable structure could be formed 
under the present experimental conditions, and thus, the vibrational modes from all of these 
structures can contribute to the peaks, causing them to broaden. Here, we briefly comment on the 
short atomic chains. As discussed before, the vibrational spectra suggest that electrons can be 
transmitted through the Pd rather than the hydrogen for contacts with conductance above 0.4 
$G_{0}$ (short Pd chain), which agreed with the theoretical calculation results.

The conductance of the Pd atomic contacts with similar atomic configurations has been 
investigated using first-principle molecular dynamics simulations \cite{18}. The conductance 
values of short and long Pd chains are reported as 1 $G_{0}$ and 0.5 $G_{0}$, respectively. It was 
found that the conductance of the short Pd chain is smaller than that of the long Pd chain. Our 
experimental results show atomic chain conductance decreased with chain length, as shown in Fig. 
~\ref{fig1}(a) and Fig.~\ref{fig2}(c), which is in agreement with the results of theoretical 
calculations. The smaller conductance values obtained in the experimental results can be explained 
as arising from the adsorption or incorporation of hydrogen onto the stems of the Pd electrodes 
since hydrogen molecules may dissociate and migrate into the bulk Pd. The theoretical calculations 
show that the conductances of the hydrogen-adsorbed and hydrogen-incorporated Pd contacts 
decreased by 30$\sim$60$\%$ depending on the amount of the hydrogen in the Pd electrodes 
\cite{19}. Our experimental and calculated results combined with previous theoretical simulations 
\cite{18} revealed that the atomic chain of two Pd atoms in contact with the hydrogen bridge can 
be formed when a Pd contact is stretched in a H$_{2}$ atmosphere while clean contacts do not, in 
fact, form an atomic chain.

\section{CONCLUSIONS}
\label{sec5}

We have investigated the atomic configuration and conductance of Pd atomic contacts in a 
H$_{2}$ atmosphere. The formation of Pd atomic chains was demonstrates using the length 
histogram of the last conductance plateau. Vibrational spectroscopy and the results of theoretical 
calculations show the formation of highly conductive short hydrogen-adsorbed Pd chain and low 
conductive long hydrogen incorporated Pd contacts.

\section{ACKNOWLEDGMENTS}
This work was supported by a Grant-in-Aid for Scientific Research on Priority Areas "Electron 
transport through a linked molecule in nano-scale", Effective Utilization of Elements 
"Nano-Hybridized Precious-Metal-Free Catalysts for Chemical Energy Conversion", and the Global 
COE Program (No. B01) from MEXT.

\newpage

\begin{references}
\bibitem{2}	A. Delin, E. Tosatti, R. Weht. Phys. Rev. Lett. {\bf 92}, 057201 (2004).
\bibitem{3}	N. Agrait, A.L. Yeyati, J.M. van Ruitenbeek, Physics Reports {\bf 377}, 81 (2003).
\bibitem{4}	H. Ohnishi, Y. Kondo, and K. Takayanagi, Nature {\bf 395}, 780 (1998).
\bibitem{5}	A.I. Yanson, G.R. Bollinger, H.E. van den Brom, N. Agrait, and J.M. van Ruitenbeek, Nature 
{\bf 395}, 783 (1998).
\bibitem{6}	R. H. M. Smit, C. Untiedt, A. I. Yanson, and J. M. van Ruitenbeek, Phys. Rev. Lett. {\bf 87}, 266102 
(2001).
\bibitem{7}	W H A Thijssen, D. Marjenburgh, R H Bremmer, J. M. van Ruitenbeek, Phys. Rev. Lett.  {\bf 96}, 
026806 (2006).
\bibitem{8}	M. Kiguchi and K. Murakoshi, Appl. Phys. Lett. {\bf 88}, 253112 (2006).
\bibitem{9}	Sz. Csonka, A. Halbritter, G. Mihaly, O.I.Shklyarevskii, S. Speller, H.van Kempen, Phys. Rev. 
Lett. {\bf 93}, 016802 (2004).
\bibitem{10}	R. H. M. Smit, Y. Noat, C. Untiedt, N. D. Lang, M. C. van Hemert and J. M. van Ruitenbeek, 
Nature {\bf 419}, 906 (2002).
\bibitem{11}	M. Kiguchi, O. Tal, S.Wohlthat, F. Pauly, M. Krieger, D. Djukic, J.C. Cuevas, and J.M. van 
Ruitenbeek, Phys. Rev. Lett. {\bf 101}, 046801 (2008).
\bibitem{12}	T. Shimazaki, Y. Asai, Phys. Rev. B {\bf 77}, 115428 (2008).
\bibitem{13}	M. Paulsson, T. Frederiksen, H. Ueba, N. Lorente, and M. Brandbyge, Phys. Rev. Lett. {\bf 100}, 
226604 (2008).
\bibitem{14}	Gaussian03, Revision C.02, M. J. Frisch, et al. Gaussian, Inc.,Wallingford CT, 2004
\bibitem{15}	A.D.Becke, J. Chem. Phys. {\bf 98}, 5648 (1993).
\bibitem{16}	D. Andrae, U. Hausermann, M. Dolg, H. Stoll, H. Preus, Theor. Chim. Acta {\bf 77}, 123 (1990).
\bibitem{17}	T.H. Dunning, Jr. J. Chem. Phys. {\bf 90}, 1007 (1989).
\bibitem{18}	B. Pieczyrak, C Gonzalez, P Jelnek, R. Perez, J. Ortega and F. Flores, Nanotechnology {\bf 19}, 
335711 (2008)
\bibitem{19}	X. Wu, Q. Li, and J. Yang, Phys. Rev. B {\bf 72}, 115438 (2005).




\end{references}
\end{document}